\documentstyle[twocolumn,floats,aps,prl,epsfig]{revtex}
\tighten
\begin{document}
\draft
\title{The Equation of State for Cold and Dense Strongly Interacting Matter
\thanks{Presented at the International Conference on Theoretical Physics 
(TH-2002), Paris, UNESCO, 22-27 July 2002.}\cite{th-proc}}
\author{E. S. Fraga$^{a,b}$, Y. Hatta$^{c,d}$, R. D. Pisarski$^e$, and J. 
Schaffner-Bielich$^f$} 
\address{$^a$Laboratoire de Physique Th\'eorique, 
Universit\'e Paris XI, 
B\^atiment 210, 91405 Orsay Cedex, France\\
$^b$Instituto de F\'\i sica, Universidade Federal do 
Rio de Janeiro, 
C.P. 68528, Rio de Janeiro, 21941-972 RJ, Brazil\\
$^c$Department of Physics, Kyoto University, Kyoto 606-8502, Japan\\
$^d$The Institute of Physical and Chemical Research (RIKEN), 
Wako, Saitama  351-0198, Japan\\
$^e$Department of Physics, Brookhaven National Laboratory, 
Upton, NY 11973-5000, USA\\
$^f$RIKEN BNL Research Center, Brookhaven National Laboratory, 
Upton, NY 11973-5000, USA}
\maketitle   


\begin{abstract}
We discuss recent results for the equation of state for cold and dense 
strongly interacting matter. We consider the extreme cases of very high 
densities, where weak-coupling approaches may in principle give reasonable 
results, and very low densities, where we use the framework of heavy-baryon 
chiral perturbation theory. We also speculate on the nature of the chiral 
transition and present possible astrophysical implications.
\end{abstract}


\bigskip

During the last decade, the investigation of strongly interacting 
matter under extreme 
conditions of temperature and density has attracted an increasing interest. In 
particular, the new data that started to emerge from the high-energy heavy ion 
collisions at RHIC-BNL, together with an impressive progress achieved by 
finite-temperature lattice simulations of Quantum Chromodynamics (QCD), provide 
some guidance and several challenges for theorists. All this drama takes place 
in the region of nonzero temperature and very small densities of the phase 
diagram of QCD (see Fig. 1). On the other hand, precise astrophysical data appear 
as a new channel to probe strongly interacting matter at very large densities. 
Compact objects, such as neutron stars, whose interior might be dense enough to 
accomodate deconfined quark matter, may impose strong constraints 
on the equation of state for QCD at high densities and low temperatures. Moreover, 
the first results from the lattice at nonzero values for the quark chemical 
potential, 
$\mu$, finally start to appear \cite{fodor,Allton:2002zi} (see Fig.2). 
Nevertheless, there is still a long way to 
go before they can achieve the high degree of accuracy obtained when $\mu=0$. As they 
improve, they will certainly bring important contributions for a better understanding 
of the phase structure of QCD.

\begin{figure}[hbt]
\centerline{\epsfig{file=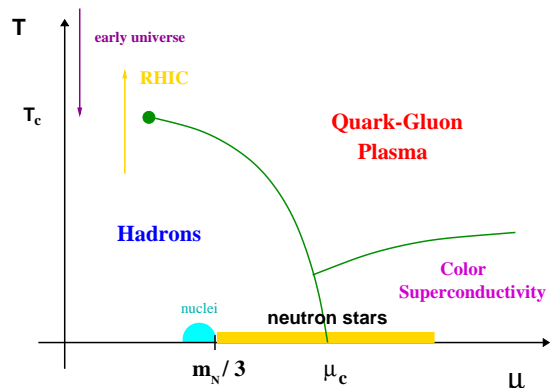,width=0.4\textwidth}}
\caption{Cartoon of the QCD phase diagram.}
\end{figure}

Recently, there was some progress in the study of the equation of state for QCD 
at zero temperature and nonzero chemical potential, especially in the regime of 
very high densities. In this region of the phase diagram, several different 
approaches, which make use of the fact that the strong coupling, $\alpha_s$, 
is relatively weak there, seem to point in the same direction and obtain a 
reasonable agreement for the pressure. Here we will discuss some of these results 
and how they compare. Then we will jump to the opposite regime in density, namely 
the case of densities much smaller than the saturation density of nuclear 
matter, $n_0$, and present some interesting results provided by heavy-baryon 
chiral perturbation theory. Of course, the region which is most interesting for 
the phenomenology of compact stars lies between the two extrema. There, both 
techniques fail completely. However, by using results obtained within some toy models, 
we will speculate on some possible scenarios for the nature of the chiral transition.

\begin{figure}[hbt]
\centerline{\epsfig{file=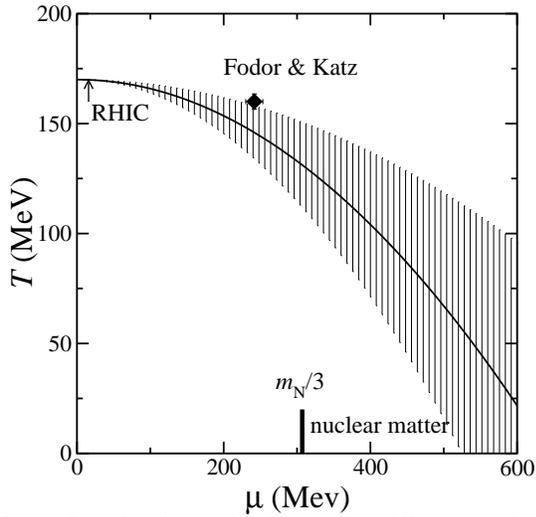,width=0.39\textwidth}}
\caption{Results from the lattice by Allton {\it et al.}, 2002. 
(Point labelled Fodor $\&$ Katz from Fodor and Katz, 2002.)}
\end{figure}

Let us first consider the case of cold and very dense strongly interacting matter. 
For high enough values of the quark chemical potential, there should be a quark 
phase due to asymptotic freedom. In this regime of densities, one is in principle 
allowed to use perturbative QCD techniques \cite{larry,baluni,fps1}, 
which may be enriched by resummation methods and quasiparticle model 
descriptions \cite{tony,andersen,andre}, to evaluate the 
thermodynamic potential of a plasma of massless quarks and gluons. Different 
approaches seem to agree rasonably well for $\mu >> 1~$GeV, and point in 
the same direction even for $\mu \sim 1~$GeV and smaller, where we are 
clearly pushing perturbative QCD far below its region of applicability. 
This is illustrated by Figs. 3--5. However, at some point between 
$\mu\approx 313~$MeV,  and $\mu\approx 1~$GeV, one has to match the equation 
of state for quark matter onto that for hadrons.

\begin{figure}[hbt]
\centerline{\epsfig{file=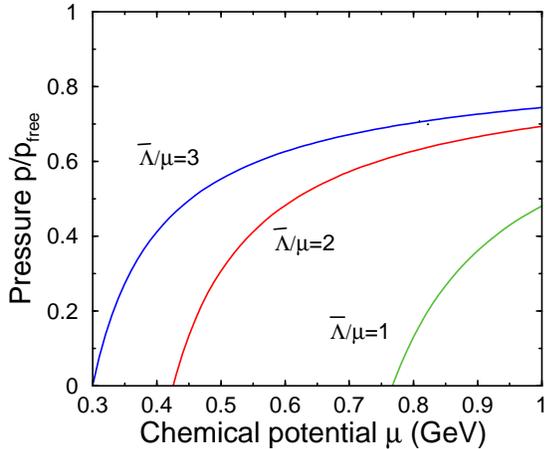,width=0.4\textwidth}}
\caption{Pressure in units of the free gas pressure as a function 
of the quark chemical potential from finite-density perturbative QCD 
(by Fraga, Pisarski and Schaffner-Bielich, 2001).}
\end{figure}
\begin{figure}[hbt]
\centerline{\epsfig{file=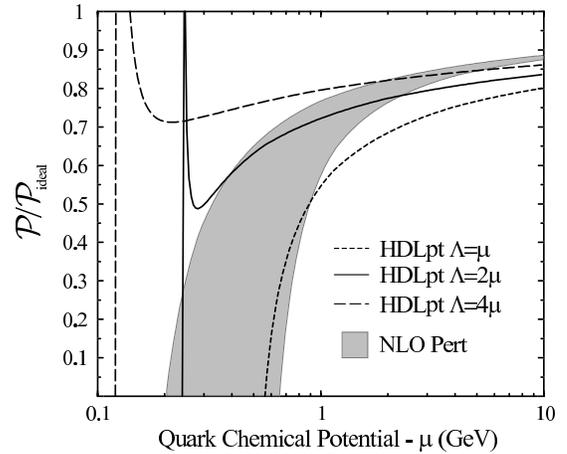,width=0.4\textwidth}}
\caption{Pressure in units of the free gas pressure as a function 
of the quark chemical potential from HDL perturbative  QCD 
(by Andersen and Strickland, 2002).}
\end{figure}

As we argued in \cite{fps1,fps2}, depending on the nature of the chiral 
transition there might be important consequences for the phenomenology of 
compact stars. For instance, in the case of a strong first-order chiral 
transition (see Fig. 6), a new stable branch may appear in the mass-radius 
diagram for hybrid neutron stars, representing a new class of compact stars 
(see Fig. 7). On the other hand, for a smooth transition, or a crossover, one 
finds only the usual branch, generally associated with pulsar data. 
This is an important issue for the ongoing debate on the radius measurement 
of the isolated neutron star candidate RX J1856.5-3754, which might be a 
quark star \cite{drake}. For details, see Refs. \cite{fps1} and \cite{fps2}.

\begin{figure}[hbt]
\centerline{\epsfig{file=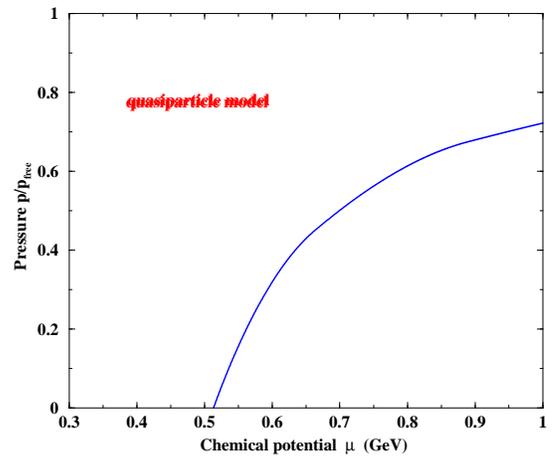,width=0.4\textwidth}}
\caption{Pressure in units of the free gas pressure as a function 
of the quark chemical potential from a quasiparticle model 
(by Peshier, 2002).}
\end{figure}
\begin{figure}[hbt]
\centerline{\epsfig{file=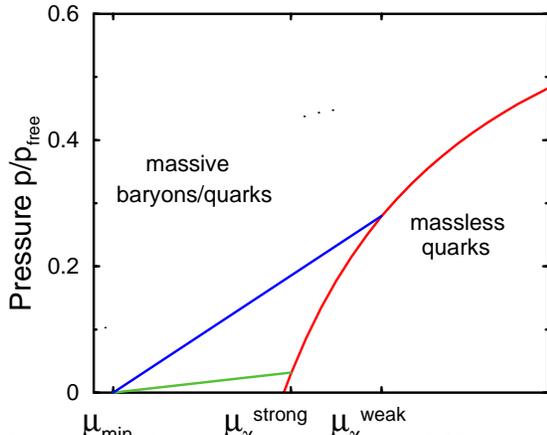,width=0.4\textwidth}}
\caption{Sketch for the matching of the pQCD equation of state onto that of
hadrons.}
\end{figure}
\begin{figure}[hbt]
\centerline{\epsfig{file=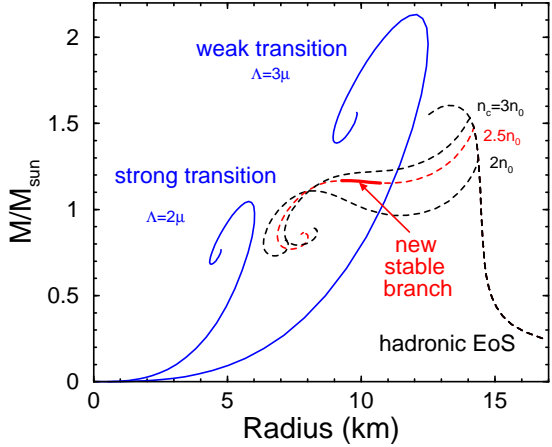,width=0.4\textwidth}}
\caption{Mass-radius diagram (for details, see Fraga, Pisarski and 
Schaffnner-Bielich, 2002).}
\end{figure}

For pure neutron (asymmetric) matter, which will play the role of hadrons 
at ``low'' density here, Akmal, Pandharipande and Ravenhall \cite{akmal} 
have found that to a very good approximation we have, up to $\sim 2 n_0$, 
the following energy per baryon: 
\begin{equation}
\frac{E}{A} -m_N = 
\frac{\epsilon}{n} - m_N \approx 15 \; {\rm MeV} 
\left(\frac{n}{n_0}\right),
\end{equation}
which is approximately linear in the baryon density, $n$. Here, 
$n_0 \sim 0.16$ baryons/fm$^3$ is the saturation density for 
nuclear matter. From this relation, we can extract the pressure:
\begin{equation}
\frac{p_{hadron}}{p_{free}} = \frac{n^2}{p_{free}} 
\frac{\partial}{\partial n} 
\left(\frac{\epsilon}{n} \right) \approx  0.04  
\; \left(\frac{n}{n_0}\right)^2 \, .
\end{equation}
From these results, we see that: (i) even at ``low'' densities we have a  
{\it highly} nonideal Fermi liquid, since free fermions would 
give $\epsilon/n - m \sim n^{2/3}$; (ii) energies are very small on hadronic 
scales (if we take $f_{\pi}$, as a ``natural scale''); (iii) energies are 
small not only for nuclear matter (nonzero binding energy),
but even for pure neutron matter (unbound). Then, this 
might be a generic property of baryons interacting
with pions, etc., and not due to any special tuning.

In order to investigate the pion-nucleon interaction at nonzero, but low, 
density, we considered the following chiral Lagrangian \cite{fhps1} 
(see also \cite{savage+wise} and \cite{kaiser})
\begin{eqnarray}
{\cal L}&=&\overline{\psi}_i\left[ 
i /\!\!\!\partial - m_N + \mu\gamma^0 - 
\frac{g_A}{2f_{\pi}}\gamma_5\gamma^{\mu}\vec{\tau}\cdot 
(\partial_{\mu}\vec{\pi}) \right] \psi_i  \nonumber \\
&+& {\cal L}_{\pi}^{0} + 
\left[ \stackrel{\mbox{other meson}}{\mbox{terms}}  \right] + 
\left[ \stackrel{\mbox{higher-order}}{\mbox{terms}}  \right] \, ,
\end{eqnarray}
where ${\cal L}_{\pi}^{0}$ is the free Lagrangian for the pions, $\vec{\pi}$, 
$\psi_i$ represent nucleons (in $n_s$ species), and $\mu$ is the chemical 
potential for the nucleons. From the Goldberger-Treiman relation we have:
\begin{equation}
g_A=\left( \frac{f_{\pi}}{m_N}\right) g_{\pi NN}
\end{equation}
and, from the Particle Data Group, $m_N=939 \,\mbox{MeV}$, 
$m_{\pi}=135 \,\mbox{MeV}$, $f_{\pi}=130 \,\mbox{MeV}$, and 
$g_{\pi NN}=13.1 \, (g_A=1.81)$

The goal is to compute the nucleon and the pion one-loop self-energy 
corrections due to the medium up to lowest order in the nucleon density 
(only nonzero $\mu$ contributions) by using the technique of heavy-baryon 
chiral perturbation theory. Therefore, we adopt a non-relativistic 
approximation:
\begin{equation}
\omega_p \approx m_N + \frac{\vec{p}\, ^2}{2m_N}+\cdots
\end{equation}
\begin{equation}
\mu \approx m_N + \frac{p_f^2}{2m_N}+\cdots
\end{equation}
Moreover, we assume that external legs are near the mass-shell 
$(p_0 + \mu)^2 - \omega_p^2 \approx 0$, that pions are 
dilute, that we have  small values of Fermi momentum, and consider 
only leading order in $(p_f/m_N)$, $(p_f/f_{\pi})$, etc.

\begin{figure}[hbt]
\centerline{\psfig{file=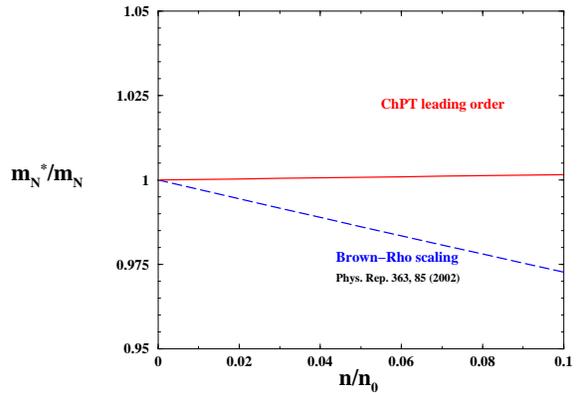,width=0.35\textwidth}}
\caption{Effective nucleon mass as a function of baryon density, 
$n$.}
\end{figure}

One-loop calculations within this framework provide the 
following result for $f_{\pi}$: as the Fermi momentum, $p_f$, 
increases -- restoring chiral symmetry -- $f_{\pi}$ must go down. 
Indeed, from chiral perturbation theory, we obtain  
($a\equiv (g_A^2/48\pi^2)$):
\begin{eqnarray}
\frac{f_{\pi}(p_f)}{f_{\pi}} &=& 1 - 
a\, \frac{p_f^3}{m_N f_{\pi}^2} + \cdots=\nonumber\\
&=& 1 - (15/939)\, (n/n_0)+ \cdots
\end{eqnarray}
Then, from Brown-Rho scaling one would expect that all quantities 
should scale in a uniform fashion. However, we obtain the following 
result for the nucleon mass:
\begin{eqnarray}
&&\frac{m_N(p_f)}{m_N}= 1+ a\, \frac{p_f^3}{m_N f_\pi^2} -\nonumber\\
&-&\frac{a}{8}\,\frac{m_\pi^2 p_f}{m_N f_\pi^2} 
\left\{ 1 + \left(\frac{m_\pi^2 +p_f^2-p^2}{4p_f^2}\right)
\log\left[ \frac{m_\pi^2+(p_f+p)^2}{m_\pi^2+(p_f-p)^2} \right]
\right\}
\end{eqnarray}

Therefore, a simple and clean chiral perturbation theory calculation 
seems to imply that, for very low densities, although $f_{\pi}$ goes down as 
we increase the Fermi momentum, $m_N$ goes up, violating the Brown-Rho 
scaling hypothesis \cite{brown-rho}. The behavior of the effective mass 
as a function of density, as predicted by chiral perturbation theory and 
by the Brown-Rho scaling is illustrated in Fig. 8. Details of the 
calculations and a more complete analysis can be found in \cite{fhps1}. 
On the other hand, as discussed in \cite{fhps1}, contact terms cannot be 
neglected and can drastically modify the results for the mass shifts.

In order to be able to make clear predictions for the phenomenology 
of compact stars, and to have a better understanding of this region 
of the QCD phase diagram, one has to find a way to describe the 
intermediate regime of densities in the equation of state, where 
perturbative calculations do not work. The study of effective field 
theory models might bring some insight to this problem \cite{progress}.

E.S.F. thanks A. Peshier and A. Rebhan for 
kindly providing some of their results for the equation of state and 
for elucidating discussions. We thank E. Kolomeitsev, M. Savage, and 
U. van Kolck for discussions. 
E.S.F. is partially supported by CAPES, CNPq, FAPERJ and FUJB/UFRJ. 
R.D.P. thanks the U.S. Department of Energy for their support under Contract
No. DE-AC02-98CH10886.



\end{document}